\documentclass[12pt]{iopart}
\usepackage{iopams} 
\input epsf
\expandafter\let\csname equation*\endcsname\relax
 \expandafter\let\csname endequation*\endcsname\relax
 \usepackage[fleqn]{amsmath}
\usepackage{amsfonts}
\usepackage{amssymb}
\usepackage{amsopn}
\usepackage{epsfig}
\usepackage{graphics,psfrag,rotating}
\usepackage{graphicx}
\usepackage{dcolumn}
\usepackage{bm}
\usepackage{epstopdf}
\usepackage{color}
\usepackage[usenames,dvipsnames,svgnames]{xcolor}
\usepackage[colorlinks=true,
      linkcolor=red,
      urlcolor=gray,
      citecolor=blue]{hyperref}
 \usepackage{hyperref}
 \usepackage[]{cite}

\def \D {\tilde{\nabla}}

\def\ga{\gamma}

\def\3nab{\tilde{\nabla}}

\def\tl{\tilde}
\def\hsp5{\hspace{5mm}}
\newcommand{\sfrac}[2]{{\textstyle{#1\over#2}}}
\def\case#1/#2{\textstyle\frac{#1}{#2}}

\def\ber {\begin{eqnarray}}
\def\eer {\end{eqnarray}}
\def\bea {\begin{eqnarray}}
\def\eea {\end{eqnarray}}

\def\bc {\begin{center}}
\def\ec {\end{center}}
\def\case#1/#2{\frac{#1}{#2}}

\newcommand{\bw}{\begin{widetext}}
\newcommand{\ew}{\end{widetext}}
\newcommand{\nn}{\nonumber\\}

\newcommand{\be}{\begin{equation}}
\newcommand{\bse}{\begin{subequation}}
\newcommand{\ese}{\end{subequation}}
\newcommand{\ee}{\end{equation}}
\newcommand{\eei}{\end{eqnarray}\indent\indent}
\newcommand{\ba}{\begin{array}}
\newcommand{\ea}{\end{array}}
\newcommand{\bal}{\begin{eqnarray}}
\newcommand{\eal}{\end{eqnarray}}

\newcommand{\Ta}{\Theta}
\newcommand{\al}{\alpha}

\def\case#1/#2{\textstyle\frac{#1}{#2} }
\newcommand{\nb}{\nabla}

\newcommand{\bver}{\begin{verbatim}}
\newcommand{\ever}{\end{verbatim}}

\begin{document}
\title{Chaplygin-gas Solutions of $f(R)$ Gravity}
\author{Maye Elmardi$^{1,2}$, Amare Abebe$^{3}$ and Abiy Tekola$^{4}$}
\address{$^1$ Astrophysics, Cosmology and Gravity Centre, University of Cape Town, Rondebosch, 7701, South Africa}
\address{$^{2}$ Department of Mathematics and Applied Mathematics, University of Cape Town, Rondebosch, 7701, South Africa}
\address{$^{3}$ Department of Physics, North-West University, Mafikeng 2735, South Africa}
\address{$^{4}$ Las Cumbres Observatory Global Telescope Network, Goleta, CA 93117, USA}
\date{\today}

\begin{abstract}

We explore  exact  $f(R)$ gravity solutions that mimic Chaplygin-gas inspired $\Lambda$CDM cosmology. Starting with the original, generalized and modified Chaplygin gas equations of state, we reconstruct the forms of $f(R)$ Lagrangians. The resulting solutions are generally quadratic in the Ricci scalar, but have appropriate $\Lambda$CDM solutions in limiting cases. These solutions, given appropriate initial conditions, can be potential candidates for scalar field-driven  early universe expansion (inflation) and  dark energy-driven late-time cosmic acceleration.
\end{abstract}
 
 \maketitle


\section{Introduction}
Explaining the not-yet-so-fully-understood accelerated expansion phases of the early and current stages of the Universe as well as the flatness in the rotational curves of spiral galaxies have certainly become in the top of to-do lists in modern cosmology and astrophysics.
As a result, current research into the nature of so-called  dark energy and dark matter has led to reconsiderations of the gravitational  physics (General Relativity, GR) and cosmic fluid (baryonic matter) models underlying the standard cosmological model.

Among a large pool of alternative theories put forward to explain early-universe inflation \cite{staro80} and late-time cosmic acceleration \cite{carroll04,magnano87,nojiri03,sotiriou07,sriva08, capoz06,nojir07} are {\it $f(R)$ theories} of gravity (see Refs. \cite{de2010,clifton12,capo11,sotiriou10, faraoni08, bamba14} for more detailed reviews). Such theories were first speculated on by Buchdahl \cite{buchdal70} but recent interest has centered on their potential candidacy  as possible infrared (IR) and ultraviolet (UV) completions of GR  (see, e.g., \cite{modesto12, modesto14, biswas12, noj07,noj11,nojir11}). More recently, these gravitational alternative theories have found  cosmological applications in, {\it inter alia}, the dynamical study of homogeneous cosmological models \cite{rippl96,nojiri06,bam8,bam9,amendola07,noj2007,CDCT05,capo6,carloni2009, leach06, cap06,goheer2009power,nojiri09,dunsby10,abebe14a, elmardi15,abebe15b} and the linear growth of large-scale structures \cite{bean07,song07, abebe12, carloni08,ananda09, abebe13, ananda08, abebe14b, dombriz08}.

For quite sometime now, the Chaplygin gas \cite{chap1904} has been considered as another alternative cosmological model of a Friedmann-Lema\^itre-Robertson-Walker (FLRW) universe with an exotic perfect-fluid  equation of state \cite{gorini2003,bento02,bento03,bilic02,deb04,dev03}. The model provides a cosmic expansion history consistent with a universe that transitions from a decelerating matter-dominated phase to a late-time accelerated one \cite{kamen01,gorini03,bilic02}. As a result, there have been several efforts to come up with a unified description of both dark matter and dark energy as manifestations of a single cosmic fluid that is the Chaplygin gas \cite{bilic02,fabris02,bean03, bouh15}.

The main objective of  this work is to come up with a unified picture of both alternative cosmological scenarios discussed above. We study models of $f(R)$ gravity which, when we impose the Chaplygin gas equations of state (EoS) to their effective pressure and energy density, produce viable exact solutions that reduce to the $\Lambda$CDM  scenario in the approximate cosmological limits.

The rest of this paper is organised as follows: In Sec. \eqref{efess} we give a brief overview of $f(R)$  gravitational models and their corresponding field equations, and describe the background cosmic evolution under consideration. We give a brief description of the Chaplygin gas model and its generalized and modified EoS in Sec. \eqref{chapl}. By comparing  these EoS with the EoS for $f(R)$ models under certain assumptions, we deduce  the form of the corresponding $f(R)$ Lagrangians. 
Finally in Sec. \eqref{concsec} we discuss the results and give our conclusions.


Natural units ($\hbar=c=k_{B}=8\pi G=1$)
will be used throughout this paper, and Latin indices run from 0 to 3.
The symbols $\nabla$, $\D$ and the overdot $^{.}$ represent the usual covariant derivative, the spatial covariant derivative, and differentiation with respect to cosmic time. We use the
$(-+++)$ spacetime signature and the Riemann tensor is defined by
\begin{eqnarray}
R^{a}_{bcd}=\Gamma^a_{bd,c}-\Gamma^a_{bc,d}+ \Gamma^e_{bd}\Gamma^a_{ce}-
\Gamma^f_{bc}\Gamma^a_{df}\;,
\end{eqnarray}
where the $\Gamma^a_{bd}$ are the Christoffel symbols (i.e., symmetric in
the lower indices) defined by
\begin{equation}
\Gamma^a_{bd}=\frac{1}{2}g^{ae}
\left(g_{be,d}+g_{ed,b}-g_{bd,e}\right)\;.
\end{equation}
The Ricci tensor is obtained by contracting the {\em first} and the
{\em third} indices of the Riemann tensor:
\begin{equation}\label{Ricci}
R_{ab}=g^{cd}R_{cadb}\;.
\end{equation}
Unless otherwise stated, primes $^{'}$ etc are shorthands for derivatives with respect to the Ricci scalar
\be
R=R^{a}{}_{a}\;
\ee
and $f$ is used as a shorthand for $f(R)$.
\section{$f(R)$ Models}\label{efess}
The  action for $f(R)$ gravity can be given as
\begin{equation}\label{lagfR}
\mathcal{A}=\int {\rm d}^4 x \sqrt{-g}\left[\frac12 f(R)+{\cal L}_{m}\right]\;,
\end{equation}
where  $f(R)$ is a general differentiable function of the Ricci scalar and $\mathcal{L}_m$ corresponds to 
the matter Lagrangian. 
For models of gravitation with such action,  applying the variational principle with respect to the metric $g_{ab}$ results in the generalized field equations
\be\label{efes}
G_{ab}=T^m_{ab}+T^{R}_{ab}\equiv T_{ab}\;,
\ee
where $T_{ab}$ is the total energy momentum tensor (EMT) and is conserved by virtue of the Einstein tensor $G_{ab}\equiv R_{ab}-\frac{1}{2}Rg_{ab}$ being a covariantly conserved quantity. Here $T^m_{ab}$ is the EMT of standard matter and  \be\label{emtf}  T^{R}_{ab}=\frac{1}{2}g_{ab}\left(f-Rf'\right)+\nb_{b}\nb_{a}f'-g_{ab}\nb_{c}\nb^{c}f' +\left(1-f'\right)G_{ab}
 \ee
 is the curvature contribution to the EMT. Since $T^m_{ab}$ is a conserved quantity, we see from Eq. \eqref{efes} that $T^R_{ab}$ should also be conserved.   In this setting, the  linearized curvature components of the thermodynamical quantities are given by
\ber
&&\label{mur}\mu_{R}\equiv T^{R}_{ab}u^{a}u^{b}=\frac{1}{2}(Rf'-f)-\Theta f'' \dot{R}+\frac{1}{3}(1-f')\Theta^2+f''\D^2R\;,\\
&&\label{pr} p_{R}\equiv \sfrac{1}{3}(T^{R}_{ab}h^{ab})=\frac{1}{6}\left(3f-Rf'-2R\right)+f''\ddot{R}+f'''\dot{R}^{2}+\frac{2}{3}\Theta f''\dot{R}\nn
&&~~~~~~~~~~~~~~~~~~~~~+\frac{1}{9}(1-f')\Theta^2 -\frac{2}{3}f''\D^2R\;,\\
&&q^{R}_{a}\equiv -T^{R}_{bc}u^{b}h^{c}_{a}=\frac{1}{3}f''\Theta \tilde{\nabla}_{a}R-f'''\dot{R}\tilde{\nabla}_{a}R -f''\tilde{\nabla}_{a}\dot{R} -\frac{2}{3}(1-f')\tl\nb_a\Theta\;,\\
&&\label{pir}\pi^{R}_{ab}\equiv T^{R}_{cd}h^{c}{}_{\langle a}h^{d}{}_{b\rangle}=f''\tilde{\nabla}_{\langle a}\tilde{\nabla}_{b\rangle}R \;,
\eer
and  the total cosmic medium is composed of standard matter and the {\it curvature fluid}, with the total thermodynamical quantities given  by
\be\label{totaltherm}
\mu\equiv\mu_{m}+\mu_{R}\;,~~~\;p\equiv p_{m}+p_{R}\;,~~~
q_{a}\equiv q^{m}_{a}+q^{R}_{a}\;,~~~\;\pi_{ab}\equiv\pi^{R}_{ab}\;.
\ee
The trace of the Einstein field equations \eqref{efes}  gives
\be\label{trace}
 3f''\ddot{R}+3\dot{R}^2f'''+3\Theta\dot{R}f''-3f''\D^2R-Rf'+2f+3p_m-\mu_m=0\;,
 \ee
 and plays an important role as a constraint relation between  matter and curvature.
 
 Analogous to the matter EoS $p_m=w_m\mu_m$, where $w_m$ is the matter EoS parameter,
 we can define the EoS for the curvature fluid as $p_R=w_R\mu_R$.
 
For FLRW spacetimes, the Ricci scalar $R$ is given by 
\be
R=2\dot{\Theta}+\frac{4}{3}\Theta^2\;,
\ee where  $\Ta$   is the cosmic expansion parameter  related to the cosmological scale factor $a(t)$ and the Hubble parameter  $H(t)$ via the equations
\be 
\Ta\equiv 3\frac{\dot{a}(t)}{a(t)}=3H(t)\;.
\ee

Solving for $\dot{\Theta}$ one can rewrite the above equation as 
\be
\dot{\Theta}=\frac{R}{2}\left(1-\frac{4}{3R}\Theta^2\right)\;.
\ee
If the Ricci scalar varies  slowly, i.e., if R is almost constant, the  solution of this ordinary differential equation (o.d.e) takes the form
\be
\Theta=\frac{1}{2}\sqrt{3R}\tanh\left[\sqrt{\frac{R}{3}}(t-t_0)\right]\;,
\ee
for  some constant of integration $t_0$ that can be taken to be the time at the commencement of the inflationary phase of expansion.  Solving for the cosmological scale factor gives
\be
a(t)=a_0\sqrt{\cosh\left[\sqrt{\frac{R}{3}}(t_0-t)\right]}\;.
\ee
For simplicity we set $t_0\simeq 0$. During steady-state exponential expansion in  a de Sitter spacetime (such as during inflation or late-time evolution),  the approximation $\dot{\Ta}\to 0$ results in
\be\label{thetacon}
R=\frac{4}{3}\Ta^2=\mbox{const}\;, a(t)=a_0e^{\frac{1}{3}\Ta t}\;,
\ee
and the matter content evolves according to
\be
\mu_m=\mu_0\left(\frac{a(t)}{a_0}\right)^{-3(1+w_m)}\;.\ee
During  such an exponentially expanding cosmic evolution phase, for an initial energy density  $\mu_0$, we see that the matter energy density decays exponentially: 
\be
\mu_m=\mu_0e^{-(1+w_m)\Ta t}\;.
\ee

\section{Chaplygin-gas Solutions}\label{chapl}
Chaplygin-gas fluid models are perfect-fluid models currently posing as candidates to unify dark energy and dark matter.  These fluid models were originally studied \cite{chap1904} in the context of aerodynamics, but only recently did they see cosmological applications \cite{bento02,gorini2003,bento03,fabris02,bouh15,mor15can}. Among the interesting features of these models is that in the FLRW framework, a smooth transition between an inflationary  phase, the matter-dominated decelerating era, and then late-time accelerated de Sitter phase of cosmic expansion can be achieved \cite{kamen01,bilic02}.
\subsection{Original and Generalized Chaplygin Gases}
In the original treatment, the negative pressure associated with the Chaplygin gas  models is related to the (positive) energy density through the EoS
\be\label{cgeos}
p=-\frac{A}{\mu^\alpha}
\ee
for positive constant $A$ and $\alpha=1$. But this was later generalized \cite{kamen01,saad14} to include $0 \leq \alpha \leq 1$. One of the  first cosmological interpretations of such a fluid model was given in \cite{barr87} where for flat universes, Eq. \eqref{cgeos} corresponds to a viscosity term that is  inversely proportional to  dust energy density.  Ever since the discovery of cosmic acceleration, however, both the original and generalized Chaplygin gas models have been extensively studied as alternatives to dark energy and/or  unified dark energy and dark matter models (see, e.g., \cite{kamen01,bento02,bilic02,bean03,mak03,set07,bouh15,mor15can}).

Now if we consider the background curvature energy density and isotropic pressure terms defined in Eqs.  \eqref{mur} and \eqref{pr} above, in the constant-curvature limiting case, we have 
\be\label{murpir}
\mu_R=\frac{1}{4}\left[R(f'+1)-2f\right]=-p_R\;.
\ee
This equation of state, with an effective EoS parameter $w_R=-1$, provides the condition for an exponential (accelerated) expansion with a constant Hubble parameter. The energy density $\mu_R$  (with its negative pressure $p_R$) remains constant and can be interpreted as playing the role of the  cosmological constant $\Lambda$.

Considering the {\it curvature fluid} as a manifestation of the Chaplygin gas with the EoS \eqref{cgeos}, we obtain
\be\label{fcgeos}
p_R=-\mu_R=-\frac{A}{\mu^\al_R}\;,
\ee
which, using Eq. \eqref{murpir}, leads to the o.d.e
\be\label{freq}
R\frac{f(R)}{dR}-2f(R)+R=4A^{\frac{1}{\al+1}}\;.
\ee
Solving  this o.d.e yields
\be\label{frsol}
f(R)=R+C_1R^2-2A^{\frac{1}{\al+1}}
\ee
for an arbitrary (integration) constant $C_1$.
We note that the $\Lambda$CDM solution $f(R)=R-2\Lambda$ is already a particular solution with  $C_1=0$ and $A=\Lambda^{\alpha+1}$. In particular, if $\alpha=0$, then $A=\Lambda$, from which, going back to Eq. \eqref{fcgeos}, one concludes $\mu_R=\Lambda$.

If we include the linearized  Laplacian term in Eqs. \eqref{mur} and \eqref{pr} and use 
 the eigenvalue $-\frac{k^2}{a^2}$ of  the covariantly defined Laplace-Beltrami operator $\D^{2}$ on (almost) FLRW spacetimes
 \be\label{pe}
 \D^{2}R=-\frac{k^2}{a^2}R\;
 \ee
 for a comoving wavenumber $k$, we obtain the second-order o.d.e
 \be\label{freql}
 B^2Rf''(R)-Rf'(R)+2f(R)-R+4A^\frac{1}{\al+1}=0\;,
 \ee
 where here we have defined
 \be
 B^2\equiv \frac{4(2+3\al)}{3(1+\al)}\frac{k^2}{a^2}\;.
 \ee
 The solution of Eq. \eqref{freql} is given, for  arbitrary constants $C_2, C_3$, by
 \be\label{frsoll}
 f(R)=R+C_2\left[R^2-2RB^2\right]+C_3\left[(R^2-2RB^2)Ei\left(1,-\frac{R}{B^2}\right)+(R-B^2)B^2e^{\frac{R}{B^2}}\right]-2A^{\frac{1}{\al+1}}\;,
 \ee
 which should reduce to the quadratic solution \eqref{frsol} for negligible values of $B^2$, i.e., for small first-order contributions to the energy density and pressure terms.
\subsection{Modified Chaplygin Gas}
Over the years, several modifications to the original and {\it generalized} Chaplygin gas models have been studied. If one considers the {\it modified Chaplygin gas} (MCG) EoS \cite{ben02,ben12,pour13,saad14,kah15,kah2015,kahya15,pour14}
\be\label{gcgeos}
p_R=\gamma\mu_R-\frac{A}{\mu^\alpha_R}\;,
\ee
then the resulting  $f(R)$ model generalizes to
\be\label{frsolll}
f(R)=R+C_4R^2-2\left(\frac{A}{\gamma+1}\right)^{\frac{1}{\al+1}}\;,
\ee
where $C_4$ is an arbitrary integration constant.

The $\Lambda$CDM solution  is a limiting case of  this generalized model when $C_4=0$ and $A=(\gamma+1)\Lambda^{\alpha+1}$. In particular, if $\alpha=0=\gamma$, then $A=\Lambda$.
 
 Following similar arguments as in the preceding subsection, if we include the linearized Laplacian contributions to  the energy density and pressure, we get Eq. \eqref{freql} generalized to
 \be\label{freqlg}
 B^2Rf''(R)-Rf'(R)+2f(R)-R+4\left(\frac{A}{\ga+1}\right)^\frac{1}{\al+1}=0\;,
 \ee
 the solution of which can be given by
 \be\label{frsollg}
 f(R)=R+C_5\left[R^2-2RB^2\right]+C_6\left[(R^2-2RB^2)Ei\left(1,-\frac{R}{B^2}\right)+(R-B^2)B^2e^{\frac{R}{B^2}}\right]-2\left(\frac{A}{\ga+1}\right)^{\frac{1}{\al+1}}\;,
 \ee 
 for an arbitrary integration constants $C_5$ and $C_6$. This solution obviously generalizes Solutions \eqref{frsol},\eqref{frsoll} and \eqref{frsolll} and should  reduce to the quadratic solution \eqref{frsol} for vanishingly small $B^2$ values.
In \cite{barr83}, it has been shown  that any quadratic Lagrangian leading to an isotropic, homogeneous cosmological model takes the form
 \be\label{ihcm}
 f(R)=R-2\Lambda-\frac{1}{6}\beta R^2\;,
 \ee
 where $\beta$ is an arbitrary, real constant. If we keep only the quadratic solution in \eqref{frsollg}, i.e., if we set $C_6=0$, the Lagrangian \eqref{ihcm} corresponds to the choice 
 \be
 C_5=-\frac{1}{6}\beta\;, B=0\;,A=(\gamma+1)\Lambda^{\alpha+1}\;.
 \ee
Another interesting fact worth pointing out here is that the condition for the existence of a maximally symmetric vacuum solution in $f(R)$ gravity\cite{barr83}
 \be
 R_0f'(R_0)=2f(R_0)
 \ee
 leads to the quadratic solution resulting in the constraint
 \be
 R_0\left(1-2C_5B^2\right)-4\left(\frac{A}{\ga+1}\right)^\frac{1}{\alpha+1}=0\;.
 \ee
 The corresponding GR de Sitter, anti-de Sitter and Minkowski solutions  $R_0=4\Lambda$  (respectively for $\Lambda>0\;, \Lambda<0$ and $\Lambda=0$) are obtained when $C_5B^2=0$ and $A=(\gamma+1)\Lambda^{\alpha+1}$.
 \subsection{Modified Generalized Chaplygin Gas}\label{mgcg}
 The so-called {\it modified generalized Chaplygin  gas} (mGCG) model  is described by a barotropic equation of state of the form \cite{bouh15,mor15can}
 \be
 p=\beta \mu-(1+\beta)\frac{A}{\mu^\alpha}\;.
 \ee
 Models of $f(R)$ gravity  that satisfy the condition \eqref{murpir}, at the sam time mimicking the mGCG, can be shown to be governed by the same equation as \eqref{freq} and admit the same solutions \eqref{frsol},  provided $\beta\neq -1$. On the other hand, if linearized Laplacian terms are included, then the corresponding differential equation in $f(R)$ generalizes to
\be\label{freqlm}
 D^2Rf''(R)-Rf'(R)+2f(R)-R+4A^\frac{1}{\al+1}=0\;,
 \ee
 where we have defined
  \be
 D^2\equiv \frac{4\left[2+3\al+3\beta(1+\al)\right]}{3(1+\al)(1+\beta)}\frac{k^2}{a^2}\;.
 \ee
 Worthy of note is that this equation and its solution
 \be\label{frsollmg}
 f(R)=R+C_7\left[R^2-2RD^2\right]+C_8\left[(R^2-2RD^2)Ei\left(1,-\frac{R}{D^2}\right)+(R-D^2)D^2e^{\frac{R}{D^2}}\right]-2\left(\frac{A}{\ga+1}\right)^{\frac{1}{\al+1}}\;,
 \ee 
 reduce to their {\it generalized} counterparts of Eqs. \eqref{freql} and \eqref{frsoll} when $\beta=0$.

\section{Discussions and Conclusion}\label{concsec}
In this paper we have explored models of $f(R)$ gravity that have Chaplygin gas equations of state. We started with the assumption that the effective energy density and isotropic pressure terms of Eqs. \eqref{mur} and \eqref{pr}, sourced by the energy-momentum tensor of the curvature fluid \eqref{emtf}, satisfy the equation of state of the Chaplygin gas, both in its original (Eq. \ref{cgeos}) and generalized (Eq. \ref{gcgeos}) forms. We then reconstructed, for time-independent Ricci scalar, the f(R) Lagrangians that allow such equations of state. When only background (zeroth-order) terms are considered for the energy density and pressure, we notice that the effective equation of state is $w_R= -1$, and hence we naturally get $f(R)$ models that are quadratic in $R$ but have limiting $\Lambda$CDM solutions.

On the other hand, keeping the first-order (Laplacian of R) terms in the energy density and pressure of the curvature fluid results in the effective energy density picking up linearized corrections to $-1$. The resulting linearized second-order o.d.es in $f(R)$ result in  more generalized solutions including exponential integral terms, of which the quadratic Lagrangians are but a subset of solutions. These solutions also  include $\Lambda $CDM as a limiting solution.

\ack The authors thank John Barrow and Behnam Pourhassan for their constructive comments.
\appendix
\section*{References}

\bibliography{bibliography}
\bibliographystyle{naturemag}

\end{document}